\documentclass[aps,prl,reprint,groupedaddress,showkeys]{revtex4-1}

\usepackage{amsmath}
\usepackage{amssymb}
\usepackage[colorlinks,linktocpage,linkcolor=blue,citecolor=blue,urlcolor=blue]{hyperref}
\usepackage{graphicx}

\usepackage{verbatim}

\renewcommand{\d}{\mathrm{d}}
\newcommand{\e}{\mathrm{e}}

\newcommand{\nl}{\notag \\ &\quad\,}
\newcommand{\nll}{\notag \\ &}

\allowdisplaybreaks

\begin{document}

\title{No classical (anti-)de Sitter solutions with O8-planes}

\author{Niccol\`{o} Cribiori$^1$ and Daniel Junghans$^2$}
\email[]{niccolo.cribiori@tuwien.ac.at}
\email[]{junghans@thphys.uni-heidelberg.de}

\affiliation{$^1$ Institute for Theoretical Physics, TU Wien, Wiedner Hauptstrasse 8-10/136, A-1040 Vienna, Austria\\$^2$ Institut f{\"{u}}r Theoretische Physik, Ruprecht-Karls-Universit{\"{a}}t Heidelberg, Philosophenweg 19, 69120 Heidelberg, Germany}

\date{\today}

\begin{abstract}
\vspace{0.35cm}
It was recently proposed that type IIA string theory may allow classical de Sitter solutions with O8-planes as the only localized sources. We show that such solutions are incompatible with the integrated supergravity equations of motion, analogously to the no-go theorem due to Maldacena and Nu\~{n}ez. We also discuss in detail divergences and discontinuities at the O8-plane positions and argue that they do not invalidate such an argument. We furthermore show that a recently proposed class of non-supersymmetric AdS solutions with O8-planes is in contrast with our results as well.\\
\end{abstract}

\maketitle

\section{I. Introduction}

Since the experimental discovery of the accelerated expansion of the universe, there has been a lively debate about whether string theory admits de Sitter (dS) vacua. Starting with \cite{Kachru:2003aw}, many constructions have been proposed, but so far they are at the level of semi-explicit scenarios rather than fully explicit solutions. 
Moreover, the ongoing search for simple, explicit models is complicated by various no-go theorems which exclude dS vacua in certain corners of the landscape \cite{Gibbons:1984kp, deWit:1986xg, Maldacena:2000mw, Hertzberg:2007wc, Steinhardt:2008nk, Caviezel:2008tf, Flauger:2008ad, Danielsson:2009ff, Caviezel:2009tu, Wrase:2010ew, VanRiet:2011yc, Green:2011cn, Gautason:2012tb, Kutasov:2015eba, Quigley:2015jia, Andriot:2016xvq, Andriot:2017jhf, Andriot:2018ept, Covi:2008ea, Shiu:2011zt, Danielsson:2012et, Junghans:2016uvg, Junghans:2016abx, Junghans:2018gdb, Banlaki:2018ayh}. This unsatisfactory state of affairs has even led some authors to conjecture that dS vacua are incompatible with quantum gravity altogether (see, e.g., \cite{Danielsson:2018ztv, Obied:2018sgi}). It is therefore important to keep looking for simple counter-examples which could falsify this claim.

In a very interesting recent paper \cite{Cordova:2018dbb}, a numerical analysis was performed suggesting that type IIA string theory may admit simple classical dS solutions with two parallel O8-planes as the only localized sources. Even ignoring possible instabilities (which were not studied in \cite{Cordova:2018dbb}), the existence of such solutions alone would already be quite amazing, given the many constraints that strongly restrict the possibility of dS solutions at the classical level. 
Indeed, all previously found classical dS solutions require intersecting O-planes and could therefore only be obtained in the smeared approximation \cite{Caviezel:2008tf, Flauger:2008ad, Danielsson:2009ff, Caviezel:2009tu, Danielsson:2010bc, Danielsson:2011au, Roupec:2018mbn, Kallosh:2018nrk, Blaback:2018hdo}. Furthermore, many of them are at small volume/large coupling \cite{Junghans:2018gdb, Banlaki:2018ayh} and known to be unstable (see \cite{Danielsson:2012et, Junghans:2016uvg, Junghans:2016abx} for an explanation of this problem and \cite{Kallosh:2018nrk, Blaback:2018hdo} for recent proposals for how to avoid it). By contrast, the dS solutions of \cite{Cordova:2018dbb} appear to work with only a minimal set of ingredients, i.e., Romans mass and two parallel O8-planes. This simplicity allowed the authors to take into account the full backreaction of the O-planes. Proving the existence of such simple dS critical points would therefore be a major step forward in the construction of fully explicit dS vacua.

However, some caution is in order, as the solutions of \cite{Cordova:2018dbb} were obtained numerically. Since the presence of the O8-planes leads to singularities in the fields and numerics with singular boundary conditions can be difficult to control, one may wonder whether the positive cosmological constant could be a numerical artifact. Indeed, the purpose of this note is to show that classical dS solutions are in fact not possible in type IIA with only O8-planes (and/or D8-branes) in the absence of further sources of higher codimension.

The argument we propose is based on integrating a specific combination of the Einstein equations and the dilaton equation over the compact space. It is a straightforward generalization of the well-known no-go theorem due to Maldacena and Nu\~{n}ez \cite{Gibbons:1984kp, deWit:1986xg, Maldacena:2000mw}, which indicates that O-planes are necessary to allow a dS$_d$ solution at the classical level. In this note, we refine such an argument and show that O8-planes (or, more generally, O8-planes and D8-branes) are not sufficient to achieve dS. Note that a similar reasoning against dS with O8-planes/D8-branes was followed earlier in \cite{Andriot:2016xvq}, where $d=4$ and a special ansatz for the dilaton was assumed.

The dS solution of \cite{Cordova:2018dbb} formally avoids the no-go theorem due to an unusual boundary condition at the O8-plane position. We show that this boundary condition is not compatible with the classical couplings of an O8-plane to the supergravity fields. We therefore interpret our result such that the dS solution of \cite{Cordova:2018dbb} is due to the presence of an unphysical source which should not be identified with an O8-plane. We also discuss in detail subtleties like divergences and discontinuities near the O8-planes and argue that the analysis we perform is not invalidated by such issues.

The same arguments can also be applied to a class of recently proposed non-supersymmetric AdS$_d$ vacua with O8-planes \cite{Cordova:2018eba}. The existence of such solutions would falsify a recent conjecture that non-supersymmetric AdS vacua are incompatible with quantum gravity \cite{Ooguri:2016pdq, Danielsson:2016mtx, Freivogel:2016qwc} and would present holographic evidence for interacting non-supersymmetric CFTs in $d-1$ dimensions. However, we find that the solutions of \cite{Cordova:2018eba} are again due to unusual boundary conditions that are not compatible with the classical couplings of an O8-plane.

\section{II. Argument}

We consider a warped compactification of type IIA string theory to $d$ dimensions with Einstein-frame metric
\begin{equation}
\d s_{10}^2 = \e^{2A} \tilde g_{\mu\nu} \d x^\mu \d x^\nu + g_{mn} \d y^m \d y^n,
\end{equation}
where the warp factor $A=A(y)$ can depend on the internal coordinates and the $(10-d)$-dimensional internal metric is kept arbitrary. We furthermore allow arbitrary dilaton $\phi=\phi(y)$, RR and NSNS fluxes and spacetime-filling localized sources of codimension 1, namely O8-planes and/or D8-branes, which may or may not be intersecting.

The classical (i.e., tree-level in $g_s$ and $\alpha^\prime$) effective action of type IIA string theory in the democratic formulation is
\begin{align}
S &= \frac{1}{2\kappa^2}\int \d^{10} x \sqrt{-g} \bigg[R -\frac{1}{2}(\partial\phi)^2-\frac{1}{2}\e^{-\phi}|H_3|^2 \nl\qquad\qquad\qquad\quad\,\,\, -\frac{1}{4}\sum_{p=0,2,4,6,8,10}\e^{\frac{5-p}{2}\phi}|F_p|^2 \bigg] \nl + \sum_i N_i\mu_8 \int \d^{9} \xi\, \e^{\frac{5}{4}\phi} \sqrt{-g} - \sum_i N_i \mu_8 \int C_9, \label{action}
\end{align}
where $F_p = (-1)^{\frac{(p-2)(p-1)}{2}}\e^{\frac{p-5}{2}\phi}\star F_{10-p}$. The last line contains the localized contributions from O8-planes and D8-branes, where $N_i=8$ for an O$8^-$-plane, $N_i=-8$ for an O$8^+$-plane and $N_i=-1$ for a D8-brane. For the corresponding anti-branes and anti-O-planes, the $C_9$ coupling receives an extra minus sign. For simplicity, we omitted further couplings of the D8-branes to $B_2$ as they play no role in the argument below.

The trace of the external components of the (trace-reversed) Einstein equations is 
\begin{align}
R^{(d)} &= \sum_{p} \frac{d(1-p)}{16} \e^{\frac{5-p}{2}\phi}|F^\text{int}_p|^2 -\frac{d}{8}\e^{-\phi}|H_3|^2 \nl
- \frac{d}{16} 2\kappa^2\mu_8 \e^{\frac{5}{4}\phi}\sum_iN_i\frac{\delta(z^i-z^i_0)}{\sqrt{g_{z^iz^i}}}. \label{einstein}
\end{align}
Here, we split all $F_p$ into those with external legs and those with purely internal parts and we dualized the external components into internal $(10-p)$-forms. Note that $F_p$ with $p<d$ are required to be internal due to the assumed maximal symmetry of the $d$-dimensional spacetime. For $d\le 3$, an extra contribution from external $H_3$ is possible, which can be shown to not change our conclusions. We denote by $z^i$ the direction transverse to the $i$th source and by $z^i_0$ the position of the source in this direction. Note that $z^i$ is in general not the same for all sources except in the special case where they are parallel.

The dilaton equation is
\begin{align}
\nabla_{10}^2 \phi &= \sum_{p} \frac{5-p}{4}\e^{\frac{5-p}{2}\phi}|F^\text{int}_p|^2 - \frac{1}{2}\e^{-\phi}|H_3|^2 \nl - \frac{5}{4}2\kappa^2\mu_8 \e^{\frac{5}{4}\phi}\sum_iN_i\frac{\delta(z^i-z^i_0)}{\sqrt{g_{z^iz^i}}}, \label{dil}
\end{align}
where $\nabla_{10}^2 = \frac{1}{\sqrt{g_{10}}}\partial_M \sqrt{g_{10}} \partial^M$. We refrain from writing down the remaining equations of motion and Bianchi identities as we will not need them for our argument.

The $d$-dimensional Ricci scalar $R^{(d)}$ can be expressed in terms of the Ricci scalar of the unwarped metric $\tilde g_{\mu\nu}$ using
\begin{equation}
R^{(d)} = \e^{-2A}\tilde R^{(d)} - \e^{-dA}\nabla_{10-d}^2 \e^{dA} \label{e1}
\end{equation}
with $\nabla_{10-d}^2 = \frac{1}{\sqrt{g_{10-d}}}\partial_m \sqrt{g_{10-d}} \partial^m$. Note also that
\begin{equation}
\nabla_{10}^2 \phi = \frac{\e^{-dA}}{\sqrt{g_{10-d}}} \partial_m \left(\e^{dA}\sqrt{g_{10-d}}\partial^m \phi\right). \label{e2}
\end{equation}

We now combine the Einstein and dilaton equations such that the source terms cancel out. Using furthermore \eqref{e1} and \eqref{e2}, we find
\begin{align}
\e^{-2A}\tilde R^{(d)} &= \frac{\e^{-dA}}{\sqrt{g_{10-d}}} \partial_m\left( \sqrt{g_{10-d}} \left( \partial^m \e^{dA} + \frac{d}{20} \e^{dA}\partial^m \phi \right)\right) \nl - \sum_{p} \frac{dp}{20} \e^{\frac{5-p}{2}\phi}|F^\text{int}_p|^2 -\frac{d}{10}\e^{-\phi}|H_3|^2. \label{r}
\end{align}
We can now multiply both sides of \eqref{r} with $\e^{dA}$ and integrate over the compact space to find
\begin{equation}
\left(\int\d^{10-d} y\sqrt{g_{10-d}} \,\e^{(d-2)A}\right) \tilde R^{(d)} \le 0. \label{r0}
\end{equation}
It follows that, for a finite warped volume, only Minkowski or AdS vacua are possible. In the special case with only Romans mass $F_0$ and no further RR or NSNS field strengths, the only possible solution is Minkowski. In order to allow classical dS in type IIA, one therefore requires localized sources with codimension 3 or higher.

\section{III. Divergences and discontinuities}

The above argument relies on the assumption that the function $\sqrt{g_{10-d}} \,\e^{(d-2)A}$ is integrable and that a total derivative on the right-hand side of \eqref{r} integrates to zero. One may wonder whether these assumptions could be wrong due to divergences or discontinuities in the fields at the O-plane positions.
Let us check that this is not the case in the simple dS solution with $d=4$ presented in \cite{Cordova:2018dbb}. This solution features one O$8^-$-plane and one O$8^+$-plane, both transverse to the same coordinate $z$. Since the only non-vanishing flux is the Romans mass, \eqref{r} simplifies to
\begin{equation}
\sqrt{g_6}\, \e^{2A}\tilde R^{(4)} = f^\prime, \label{rr}
\end{equation}
where $^\prime = \frac{\d}{\d z}$ and we write
\begin{equation}
f = \sqrt{g_{6}} \e^{-\frac{1}{5}\phi} g^{zz} \left( \e^{4A+\frac{1}{5}\phi} \right)^\prime \label{f}
\end{equation}
as a short-hand for the field combination inside of the total derivative.

For concreteness, we consider an O8$^-$-plane at $z=z_0$ as in \cite{Cordova:2018dbb}. The leading behavior of the relevant functions at small $t=z-z_0$ is
\begin{equation}
\e^{2A} \sim |t|^\frac{1}{8}, \quad\!\! \e^\phi \sim |t|^{-\frac{5}{4}}, \quad\!\! \sqrt{g_6} \sim |t|^\frac{7}{8}, \quad\!\! g_{zz} \sim |t|^\frac{9}{8}. \label{bc2}
\end{equation}
Note that this agrees with the standard behavior near an O$8^-$-plane in flat space \footnote{Substituting \eqref{bc2} back into \eqref{einstein} and integrating over the compact space naively yields a divergent contribution of the O-plane term to the $d$-dimensional energy-momentum tensor. If a consistent solution to the supergravity equations exists with the boundary conditions \eqref{bc2}, there should then be a suitable regularization procedure by which the divergent O-plane term can be shown to combine with the backreacted field strengths into a finite contribution, as required from the fact that the left-hand side of the $d$-dimensional Einstein equation is finite as well. We will proceed with the assumption that the boundary conditions \eqref{bc2} can be made sense of this way, as otherwise they should be discarded right from the start and there is nothing left to show. In any case, the divergence does not play a role for the arguments in this note as they are based on studying \eqref{r}, where all source terms cancel out.}.

Substituting \eqref{bc2} into \eqref{f} including subleading corrections, we find
\begin{align}
f &= \left( 1 + b_1|t| + b_2t^2 + \ldots \right)\times \nl\quad \left( c_0 + c_1|t| + c_2t^2+ c_3|t|^3 + \ldots \right)^\prime \nll = c_1 \left(2\Theta(t)-1\right) + (b_1c_1+2c_2)t \nl + (b_2c_1+2b_1c_2+3c_3)t|t| + \ldots, \label{rr0}
\end{align}
where $b_i$, $c_i$ are expansion coefficients and $\Theta(t)$ is the Heaviside function. Crucially, for $c_1\neq 0$, $f$ has a discontinuity at $t=0$. Taking the derivative, we obtain the right-hand side of \eqref{rr}:
\begin{align}
f^\prime &= 2c_1 \delta(t) + (b_1c_1+2c_2) \nl + 2(b_2c_1+2b_1c_2+3c_3)|t| + \ldots. \label{rr1}
\end{align}
The left-hand side of \eqref{rr} is instead
\begin{equation}
\sqrt{g_6}\,\e^{2A}\tilde R^{(4)} = d_1|t| + \ldots, \label{rr2}
\end{equation}
where $d_1$ is again some expansion coefficient.

We thus observe that the factor $\sqrt{g_6}\,\e^{2A}$ does not diverge at the O$8^-$-plane position but actually goes to zero. Using analogous arguments, one checks that $\sqrt{g_6}\,\e^{2A}$ is finite at the O$8^+$-plane. Away from the O-planes, the function must be smooth. This implies that the function is integrable and that the warped volume $\int \d^6 y \sqrt{g_6}\,\e^{2A}$ is finite as expected.

Substituting \eqref{rr1} and \eqref{rr2} into \eqref{rr}, we furthermore conclude that $c_1=c_2=0$. With \eqref{rr0}, this implies that $f$ does not have a discontinuity at the position of the O$8^-$-plane,
\begin{equation}
\Delta f = 0. \label{deltaf}
\end{equation}
An analogous argument confirms that there are no discontinuities at the O$8^+$-plane position.

Note that the functions appearing in \eqref{f} are related to the string-frame quantities in \cite{Cordova:2018dbb} by $\e^{2A} = \e^{2W-\frac{1}{2}\phi}$, $\sqrt{g_6} = \e^{-6W-\frac{3}{2}\phi+5\lambda}\sqrt{g_{M_5}}$ and $g_{zz}=\e^{-2W-\frac{1}{2}\phi}$, which implies $f \sim \e^{5\lambda-2\phi}\left(W^\prime -\frac{1}{5}\phi^\prime \right)$ with $\e^{5\lambda-2\phi}\sim t^0$. The condition \eqref{deltaf} is therefore equivalent to $\Delta W^\prime-\frac{1}{5}\Delta \phi^\prime =0$ stated in Eq.~(4a) of \cite{Cordova:2018dbb}.

Let us emphasize here that imposing continuity of $f$ at the O-plane position is not an arbitrary assumption. It follows from the fact that all delta-function sources cancel out in the combination \eqref{r} such that the term $\sim c_1\delta(t)$ is not supported in that equation. The precise way of how the delta functions appear in equations such as \eqref{einstein} and \eqref{dil} follows from the couplings of the fields in the classical action \eqref{action}. It determines, for example, the tension and the $C_9$ charge of the localized sources and can therefore not be changed if the object is to be interpreted as an O-plane in the usual manner. The boundary conditions for the first derivatives of the fields at the O-plane positions are thus fixed.

We have therefore established that $f$ is continuous everywhere on the compact space, which implies $\int \d z f^\prime = 0$. Integrating \eqref{rr} over the compact space then yields $\tilde R^{(4)}=0$, as claimed in Section II.

\begin{figure}[h!]
\includegraphics[width=.7\linewidth]{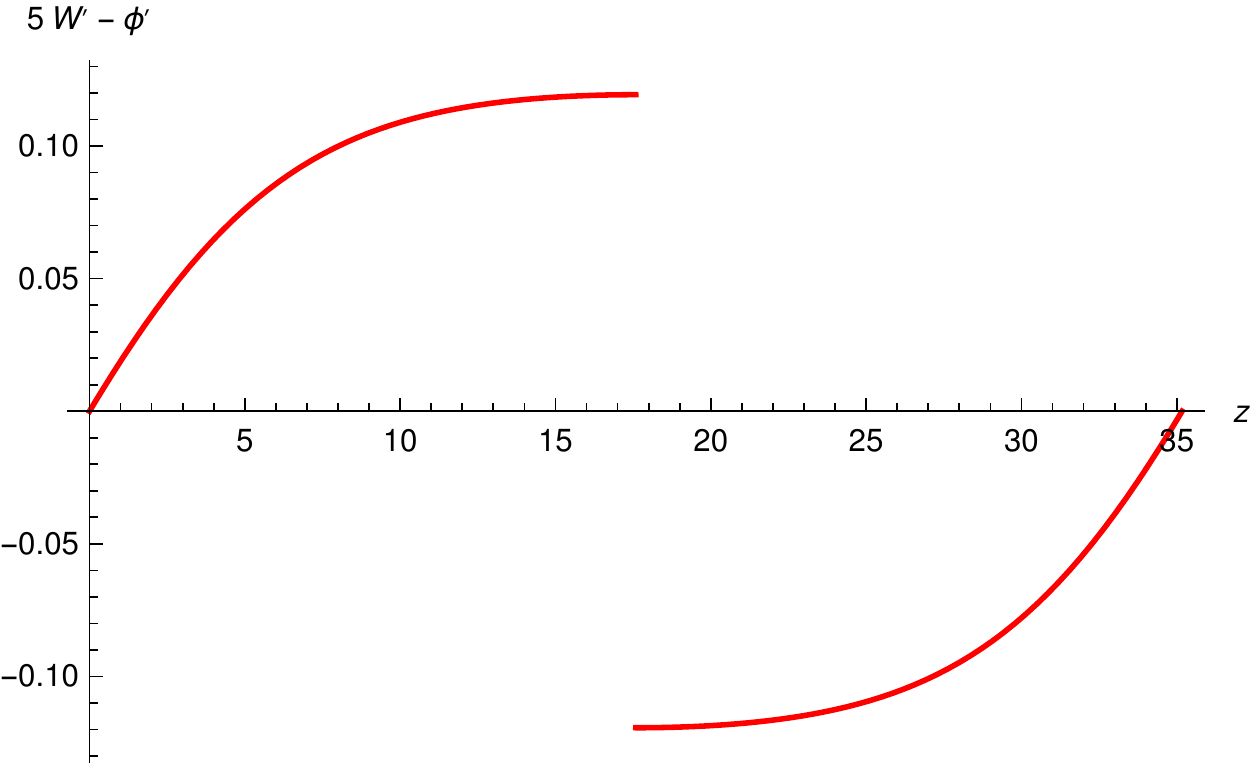}
\caption{Plot of $5W'-\phi'$ for a numerical solution to Eqs.~(2a)--(2c) in \cite{Cordova:2018dbb}. The O$8^+$-plane is located at $z=0$ while the O$8^-$ is at $z = 17.58\ldots$. The function is clearly discontinuous at this latter point.}
\label{fig1}
\end{figure}

In the numerical solution of \cite{Cordova:2018dbb}, this conclusion is formally avoided because $f$ has a discontinuity at the O$8^-$-plane (cf.~Fig.~\ref{fig1}).
Note that this is difficult to see numerically since $\Delta W^\prime$ and $\frac{1}{5}\Delta \phi^\prime$ both diverge and agree at the leading order $1/t$.
Nevertheless, there is a finite discontinuity $\Delta f \sim \Delta W^\prime-\frac{1}{5}\Delta \phi^\prime \neq 0$, incompatible with \eqref{deltaf}.

As explained above, we interpret this such that the equations of motion are not satisfied with the correct boundary conditions at the O-plane positions. In particular, we now have $c_1\neq 0$, such that an unphysical source appears on the right-hand side of \eqref{rr} and the equation is not satisfied at $t=0$.

We also observe that the non-supersymmetric AdS$_d$ solutions proposed in \cite{Cordova:2018eba} suffer from the same problem. Just like their dS cousins, these solutions have Romans mass as the only non-zero flux and should therefore be excluded by \eqref{r0}, which in this simple case becomes an equality, $\tilde R^{(d)}=0$. The relevant function inside of the total derivative of \eqref{r} is now
\begin{equation}
f_d = \sqrt{g_{10-d}}\, \e^{-\frac{d}{20}\phi} g^{zz} \left( \e^{dA+\frac{d}{20}\phi} \right)^\prime,
\end{equation}
which reduces to \eqref{f} for the case $d=4$. As before, the field behavior near the O8-planes implies that $f_d$ satisfies a continuity condition $\Delta f_d = 0$. The transverse coordinate $z$ in the examples of \cite{Cordova:2018eba} is a polar angle of an $S^{10-d}$. We therefore have $\int \d z f_d^\prime=f_d(\pi)-f_d(0)$. One checks that $f_d$ vanishes at both poles of the sphere if the fields are regular there, and we conclude that $\int \d z f_d^\prime=0$ as expected.

\begin{figure}[h!]
\includegraphics[width=.7\linewidth]{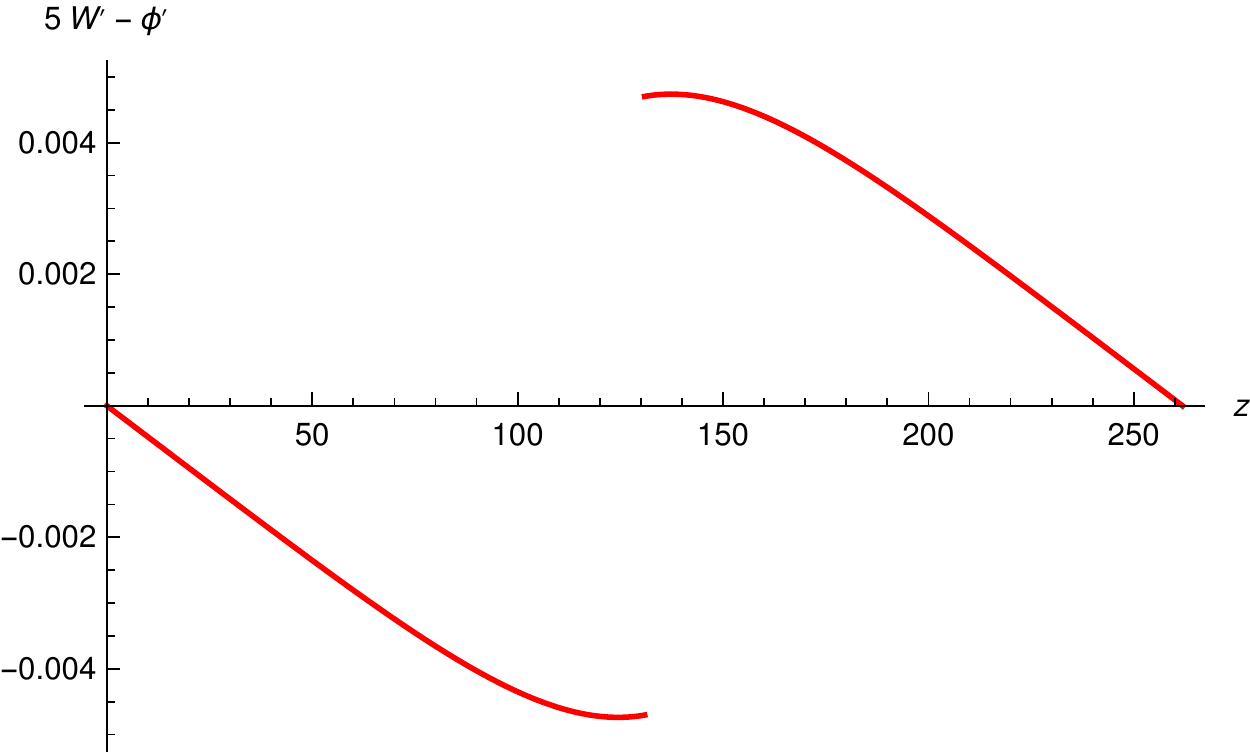}
\caption{Plot of $5W'-\phi'$ for a numerical solution to Eqs.~(2.9a)--(2.9c) in \cite{Cordova:2018eba}. Here, $z\in[0,261.84\ldots]$ is the polar angle of the $S^2$ up to a normalization factor. The O$8^-$-plane is at $z = 130.92\ldots$. The function $5W'-\phi'$ is clearly discontinuous at this point.}
\label{fig3}
\end{figure}

Considering, for example, the AdS$_8$ solution of \cite{Cordova:2018eba}, we have $f_8 \sim \e^{8W-2\phi+\lambda}\left(W^\prime - \frac{1}{5}\phi^\prime\right)$ with $\e^{8W-2\phi+\lambda}\sim t^0$ in their conventions. Hence, $\Delta f_8 = 0$ implies $\Delta W^\prime - \frac{1}{5}\Delta \phi^\prime = 0$. However, plotting this for a numerical solution to Eqs.~(2.9a)--(2.9c) in \cite{Cordova:2018eba}, we observe that the boundary condition is violated, see Fig.~\ref{fig3}. We interpret this again as the presence of an unphysical source.

Analogous arguments to those discussed above can be made for more general setups with an arbitrary distribution of O8-planes/D8-branes and, in the dS case, in the presence of arbitrary RR and NSNS field strengths. In particular, one verifies that \eqref{r} is integrable and $f_d$ is continuous in any such setup provided that the solution locally looks like an O8-plane/D8-brane in flat space very close to the sources. It would also be interesting to analyze the possible effects of discontinuities in setups with intersecting sources, where the boundary conditions may be different. Note that \eqref{r} is expected to be integrable in such cases as well since the warped volume is related to the $d$-dimensional Planck mass, which should be finite in any compactification.

\section{IV. String corrections}

So far, we were concerned with the possibility that string theory admits classical dS solutions. The basic idea \cite{Hertzberg:2007wc, Silverstein:2007ac} is to construct toy models that are as simple and explicit as possible and do not depend on any ingredients that are incompletely understood or cannot be controlled.
In particular, as we have been assuming within the present work, a classical dS solution with O8-planes should satisfy two properties:
\begin{itemize}
\item A positive cosmological constant is obtained by solving the equations of motion in the supergravity approximation, i.e., by only taking into account the terms appearing in the classical effective action \eqref{action}.
\item Stringy corrections to the classical value of the 4d cosmological constant are shown to be small such that they can self-consistently be neglected.
\end{itemize}
Crucially, the second requirement does \emph{not} mean that string corrections to the fields $\phi(y)$, $A(y)$, etc.\ are required to be small everywhere on the compact space. In fact, such a requirement would often be impossible to fulfill as string corrections typically blow up near O-planes. However, integrating the external components of the Einstein equations over the compact space, one finds that the 4d cosmological constant is only sensitive to the \emph{integrated} effect of the string corrections. Hence, even if stringy effects such as $\alpha^\prime$ and $g_s$ corrections blow up very close to the O-planes, their effect on the 4d cosmological constant may still be small if they are negligible almost everywhere on the compact space. This is usually indeed the case in the limit of large enough volumes and small string coupling. For example, $\alpha^\prime$ and $g_s$ corrections to the 4d scalar potential (and hence the 4d cosmological constant) of type IIB flux compactifications on warped Calabi-Yaus are organized in an inverse-volume and string coupling expansion, even though these compactifications feature O3-planes and O7-planes with locally large string corrections \cite{Giddings:2001yu}.

In the previous sections, we showed that, contrary to what one might conclude from \cite{Cordova:2018dbb}, the supergravity equations do not admit classical dS solutions with O8-planes (assuming the two conditions stated above). One might now argue that the supergravity equations break down close to the O8-planes such that any argument based on integrating these equations over the compact space is bound to fail. However, while it is true that string effects become relevant near the O8-planes, this is besides the point.
In fact, the intention of \eqref{r0} is not to bound the exact value of the cosmological constant for the full string-corrected solution but rather to serve as a consistency condition for the supergravity solution. It shows that the classical contribution to the cosmological constant is non-positive in general and zero for the simple setup with $F_0$ as the only flux.
Analogous integration arguments are used in the familiar setup of type IIB string theory with O3/O7-planes and 3-form fluxes, leading to the well-known fact that $\tilde R^{(4)}=0$ classically there as well \cite{Giddings:2001yu}.

In our setup with O8-planes, a positive cosmological constant can therefore only be generated by string effects such as $\alpha^\prime$ or $g_s$ corrections. Whether such string corrections have the right form to produce a dS vacuum with O8-planes was not studied in \cite{Cordova:2018dbb}. The assumption of an unusual discontinuity that lifts the solution to dS is therefore speculative at this point. Nevertheless, it is interesting to entertain the possibility that string effects could alleviate the bound \eqref{r0}. In the following, we will discuss two qualitatively different effects. The first type are corrections to the O8-plane action, which would modify the boundary conditions at the O-plane position and thus potentially introduce new discontinuities in the field derivatives. The second type are perturbative bulk corrections such as products of Riemann tensors or higher powers of RR and NSNS field strengths.

\subsection{a. String-corrected boundary conditions}

As we have seen, the dS solution of \cite{Cordova:2018dbb} stands or falls on the unusual boundary condition $\Delta f \neq 0$, which does not agree with the known couplings of an O8-plane in the last line of \eqref{action}. Nevertheless, one may speculate that stringy corrections to the O-plane action have the effect that the unusual discontinuity is sourced.

Since perturbation theory is expected to break down near the O-plane, it is difficult to guess how exactly the fields should behave there. However, a simple check we can perform is to consider a perturbative $\alpha^\prime$ or $g_s$ correction to the O-plane action. In particular, let us postulate that such a correction becomes relevant at the O-plane in such a way that the additional couplings modify the boundary conditions as desired. Our ansatz for the string-corrected O-plane action is then
\begin{equation}
\int \d^{9} \xi\, \e^{\frac{5}{4}\phi} \sqrt{-g} \to \int \d^{9} \xi \left( \e^{\frac{5}{4}\phi} \sqrt{-g} + h \right), \label{h}
\end{equation}
where $h$ is some function of the fields and their derivatives due to, e.g., an $\alpha^\prime$ or $g_s$ correction. Alternatively, we may imagine that another codimension-1 object sits on top of the O-plane and produces the extra coupling $h$. Upon varying the modified action, $h$ introduces a further delta function into \eqref{rr} such that
\begin{equation}
\sqrt{g_6}\,\e^{2A}\tilde R^{(4)} = f^\prime + C \delta(t),
\end{equation}
where $C = \frac{16\kappa^2 \mu_8}{\sqrt{-\tilde g_4}}\left(-\frac{1}{2}g^{\mu\nu}\frac{\delta h}{\delta g^{\mu\nu}} +\frac{1}{2} g^{mn}\frac{\delta h}{\delta g^{mn}}+\frac{1}{5}\frac{\delta h}{\delta \phi}\right)$. For $C\neq 0$, it may then be possible to avoid our argument against (A)dS solutions in this setup \footnote{In a previous version of this paper, we claimed that $C\neq 0$ is not sufficient to avoid our argument. This was based on the assumption that a function $\tilde f$ defined there is continuous, which need not be the case.}. We leave a detailed study of localized corrections of the form \eqref{h} to future work.

\subsection{b. Bulk corrections}

Let us finally discuss the effect of bulk $\alpha^\prime$ or $g_s$ corrections to the classical effective action. This yields additional higher-derivative terms in the equations involving, e.g., powers of the Riemann tensor and the RR and NSNS field strengths.
Let us denote such higher-derivative terms collectively by $X$, i.e., we allow that \eqref{action} is corrected such that $S \to S + \frac{1}{2\kappa^2}\int \d^{10} x X$. Eq.~\eqref{r0} thus schematically becomes
\begin{equation}
\left(\int\d^6 y\sqrt{g_6} \,\e^{2A}\right) \tilde R^{(4)} \le \int \d^6 y \sqrt{g_6}\, Y,
\end{equation}
where $\sqrt{g_6}\, Y=\frac{1}{\sqrt{-\tilde g_4}}\left(-\frac{1}{2}g^{\mu\nu}\frac{\delta X}{\delta g^{\mu\nu}} +\frac{1}{2} g^{mn}\frac{\delta X}{\delta g^{mn}}+\frac{1}{5}\frac{\delta X}{\delta \phi}\right)$.
Hence, taking into account $\alpha^\prime$ or $g_s$ corrections, the no-go against dS may in principle be violated. To our knowledge, such corrections have not been systematically analyzed in the literature in a setup with O8-planes and Romans mass (see, however, \cite{Blaback:2019zig} for a recent dS construction in M-theory). In particular, it is not known whether $\int \d^6 y \sqrt{g_6} \,Y$ can be positive in a stabilized vacuum while maintaining perturbative control. Further studies in this direction would certainly be important. In any case, such $\mathcal{O}(\alpha^\prime)$ dS vacua, if existent, should be contrasted with \emph{classical} dS solutions, where the cosmological constant is generated by the terms in \eqref{action}. What we showed in this paper is that these contributions to the cosmological constant always add up to something non-positive if only codimension-1 sources are present.
\\

\begin{acknowledgments}
We would like to thank G.~Bruno De Luca, Alessandro Tomasiello, Thomas Van Riet and Timm Wrase for very helpful correspondence.
\end{acknowledgments}

\bibliography{groups}

\end{document}